\documentstyle[prl,aps,multicol]{revtex}
\newcommand{\beq}{\begin{equation}}
\newcommand{\eeq}{\end{equation}}
\newcommand{\bqa}{\begin{eqnarray}}
\newcommand{\eqa}{\end{eqnarray}}

\def\square{\vcenter{\vbox{\hrule height.4pt
          \hbox{\vrule width.4pt height8pt
        \kern8pt\vrule width.4pt}\hrule height.4pt}}}

\begin{document}

\title{Effective Field Theories for the Casimir Effect \\at Finite Temperature}
\author{Jens O. Andersen\\
Institute for Theoretical Physics, 
University of Utrecht,\\
       Leuvenlaan 4, 3584 CE Utrecht, The Netherlands}
\date{\today}
\maketitle

\begin{abstract}
We discuss corrections to the Casimir effect at finite 
temperature
and effective field theory.
Recently, it has been shown 
that effective field theories can reproduce 
radiative corrections
to the Casimir energy calculated 
in full QED.
We apply effective field theory methods 
at finite temperature and reproduce the Casimir free energy.
We show that the 
system undergoes dimensional reduction at high temperature and 
that it can be described by an
effective three-dimensional field theory.
\end{abstract}
\setcounter{page}{1}
\vskip 0.5cm
\begin{multicols}{2}
One of the most remarkable manifestations of quantum fluctuations is the
attractive Casimir force between two parallel perfectly conducting 
plates separated by a distance $L$. It was first predicted by Casimir
in 1948~\cite{cas} and experimentally verified on a 
qualitative level 
by Sparnaay~\cite{sp}. 
Recently, experiments with much higher precision have been 
performed~\cite{lam}.

Most calculations of the Casimir effect in various geometries 
have been for free fields (See Ref.~\cite{rev} for a thorough review). 
In QED, the 
leading result for the ground state energy 
for a system of two parallel perfectly conducting plates separated by a
distance $L$,
is $E_0^{(0)}=-\pi^2/720L^3$.
The first calculation of radiative
corrections to the Casimir energy was carried out 
by Bordag, Robashick, and Wieczorek
in QED~\cite{bord}. 
In the physically interesting limit where the electron mass 
$m$ is much larger than $1/L$,
the leading radiative
correction is $E_0^{(1)}=\pi^2\alpha/2560mL^4$.

The distance $L$ between the plates introduces a new length scale in
the problem. If $m\gg1/L$ we have two vastly different scales
and one 
should be able to apply effective field theory 
methods~\cite{lepage}  to calculate
the Casimir energy. 
The first attempt to apply these 
methods
to the Casimir
problem, was made by Kong and Ravndal~\cite{finn}. Using the 
Euler-Heisenberg effective Lagrangian, 
they calculated the leading 
radiative correction. The result is 
$E_0^{(1)}=11\pi^4\alpha^2/(2^73^55^3m^4L^7)$, which disagrees 
with the result from full QED.
The application of effective field theory methods to the Casimir
problem was later questioned in Ref.~\cite{bor1}.
However, the disagreement between the two results, is not
a failure of effective field theory as such, but rather that
an incorrect effective field theory was used~\cite{f+j,mel}.
It is essential  in effective field theory calculations to first 
identify the symmetries of the effective Lagrangian, and then
impose the correct boundary conditions on the propagators before
one determines the parameters in ${\cal L}_{\rm eff}$
by matching.
Indeed, Ravndal and Thomassen~\cite{f+j}, and Melnikov~\cite{mel}
have recently
constructed effective
field theories for the electromagnetic field that could reproduce 
the leading
radiative correction to the Casimir energy in QED, if the parameters in 
the effective Lagrangian are chosen appropriately and the photon
propagator satisfies the boundary conditions. 

The zero-temperature result for the two-loop ground state energy
in~\cite{bord} was later
generalized to finite temperature by 
Scharnhorst, Robaschik, and Wieczorek~\cite{krw}.
In this paper, we consider 
the two-loop Casimir free energy
from an effective field theory point of view.
It is shown that effective field theory can reproduce the 
corrections at finite temperature in the same way as in the 
zero-temperature case. For simplicity, we consider only the
low-temperature and high-temperature limits.
In the high-temperature limit, we show that the system undergoes 
dimensional reduction and that it can be described by an
effective three-dimensional theory.

The electromagnetic field satisfies the standard metallic boundary
conditions ${\bf n}\!\times\!{\bf E}={\bf n}\!\cdot\!{\bf B}=0$, where 
$n=(0,0,1)$ is a unit vector normal to the plates.
If one introduces the four-vector $n_{\mu}=(0,0,1,0)$, we can write the
boundary conditions as $n_{\mu}\tilde{F}^{\mu\nu}=0$, where 
$\tilde{F}_{\mu\nu}=\epsilon_{\mu\nu\alpha\beta}F^{\alpha\beta}$ is 
the dual field strength~\cite{f+j}.

The implementation of the boundary conditions manifests itself in two ways.
Firstly, it modifies the contribution from the photon field to
the one-loop free energy to~\cite{bor2}:
\bqa\nonumber
{\cal F}_{0}&=&-{i\over2}d\int{d^{d+2}k\over(2\pi)^{d+2}}\log(-k^2)
\\ 
\label{1lf}
&&-{i\over2}d\int{d^{d+1}k_{\perp}\over(2\pi)^{d+1}}
\Bigg[\log(1-e^{2i\gamma L})
-\log(-\gamma^2)
\Bigg]
\;,
\eqa
where $k^{\perp}_{\mu}=(k_0,k_1,k_2)$, 
$\gamma=\sqrt{k_0^2-k_1^2-k_2^2}$, $d=2-2\epsilon$, and
$k^2=\gamma^2-k_3^2$. The first term is standard and vanishes
at zero temperature with dimensional regularization. At finite
temperature, it gives
the ideal gas contribution to the free energy, which must be kept
in order to obtain the correct high-temperature limit.
The second and third terms are due to the boundary conditions.
Secondly, the propagator is modified
due to the boundary conditions.
The propagator in Feynman gauge 
is~\cite{bor2}:
\bqa
D^{S}_{\mu\nu}(x,x^{\prime})&=&
D_{\mu\nu}(x,x^{\prime})+
\bar{D}_{\mu\nu}(x,x^{\prime})\;,
\eqa
where 
\bqa
D_{\mu\nu}(x,x^{\prime})
&=&-\int{d^{d+2}k\over(2\pi)^{d+2}}
{g_{\mu\nu}\over k^2}
e^{-ik(x-x^{\prime})}\;, \\
\nonumber
\bar{D}_{\mu\nu}(x,x^{\prime})&=&
\int{d^{d+1}k_{\perp}\over(2\pi)^{d+1}}
{P^{\perp}_{\mu\nu}\over4\gamma\sin\gamma L}
e^{-ik_{\perp}(x_{\perp}-x_{\perp}^{\prime})}\\ \nonumber
&&\times\left[
e^{-i\gamma L}\left(
e^{i\gamma|z|}e^{i\gamma|z^{\prime}|}+e^{i\gamma|z-L|}
e^{i\gamma|z^{\prime}-L|}
\right)
\right. \\
&&
\label{rtprop}
\left.\left(
-e^{i\gamma|z|}e^{i\gamma|z^{\prime}-L|}+
e^{i\gamma|z-L|}e^{i\gamma|z^{\prime}|}\right)
\right]\;.
\eqa
Here, $P_{\mu\nu}^{\perp}$ is a 
projection operator
\bqa
P_{\mu\nu}^{\perp}=
\Bigg\{\begin{array}{cc}
g_{\mu\nu}-{k_{\mu}k_{\nu}\over k^2_{\perp}}& \hspace{-1.cm}{\rm for}
\hspace{0.2cm}\mu, \nu\neq 3\\
0& \hspace{0.4cm}
{\rm for}\hspace{0.2cm}\mu=3\hspace{0.2cm}{\rm or}\hspace{0.2cm}\nu=3\;.
\end{array}
\eqa
Loop corrections to Eq.~(\ref{1lf}) are given in terms of Feynman
diagrams, but with the modified 
propagator Eq.~(\ref{rtprop}) and the standard fermion propagator~\cite{bor2}.

In the low-temperature limit $\beta\gg L$, the two-loop free energy 
reduces to~\cite{krw}
\bqa
\label{lt}
{\cal F}_{0+1}&=&-{\pi^2\over720L^3}\left[
1-{9\alpha\over32mL}
\right]
\;.
\eqa
Note that this result is in fact {\it temperature independent}. This is
due to cancellations between various temperature-dependent terms in 
this limit.

In the high-temperature limit $\beta\ll L$, the two-loop free energy 
is~\cite{krw}
\bqa\nonumber
{\cal F}_{0+1}&=&
-{\pi^2L\over45\beta^4}
\left[1-{3\alpha\over32mL}\right]
+{\zeta(3)\over2\pi\beta^3}
\\
&&\label{ht}
-{\zeta(3)\over8\pi\beta L^2}\left[
1-{3\alpha\over16mL}
\right]
\;.
\eqa
These results are not exactly those found in 
Ref.\cite{krw}. They neglected $L$-independent contributions, 
but they must be included to obtain the full result.

Melnikov~\cite{mel} considered the leading corrections to the 
Casimir energy from an effective field theory point of view.
By calculating two-loop diagrams that contribute to 
the energy density $T_{00}(z)$,
he showed that by adding 
\bqa
\delta{\cal L}={3\alpha\over32mL}
\left(-{\bf E}_z^2+{\bf B}_{\perp}^2\right)\;,
\eqa
to the Maxwell term, he reproduced the correction $E_0^{(1)}$.

More generally, one
ought to be able to construct an effective field theory
for the electromagnetic field that can be applied in the limit 
$mL\gg1$. This is done by writing down the most general Lagrangian which is
consistent with the symmetries and tune the parameters of the theory 
so that it reproduces a set of observables in full QED in the limit
$mL\gg1$. While Lorentz invariance is broken due to the boundary
conditions, gauge invariance is still a good symmetry.
The terms in ${\cal L}_{\rm eff}$ are then constructed from the
four-vector $n_{\mu}$, the field $F_{\mu\nu}$, 
its dual $\tilde{F}_{\mu\nu}$
and derivatives thereof~\cite{f+j}:
\bqa\nonumber
{\cal L}_{\rm eff}&=&
-{1\over4}F_{\mu\nu}F^{\mu\nu}+b_1\left(n_{\mu}F^{\mu\alpha}\right)
\left(n^{\nu}F_{\nu\alpha}\right) 
\\&&
\label{mell}
+\left(\partial_{\mu}\bar{\eta}\right)\left(\partial^{\mu}\eta\right)
+{\cal L}_{\rm gf}+\delta{\cal L}^{}_{\rm eff}
\;,
\eqa
where $\eta$ is the ghost field, ${\cal L}_{\rm gf}$ is a gauge-fixing
term, and 
$\delta{\cal L}_{\rm eff}$ includes all higher order operators that
satisfy the symmetries. 
It follows from 
Melnikov's result~\cite{mel} 
that the coefficient $b_1=3\alpha/32mL$.

A somewhat different approach was used by Ravndal and Thomassen~\cite{f+j}. 
They apply an effective Lagrangian 
that consists of both bulk terms and surface terms:
\bqa
\label{finnl}
{\cal L}_{\rm eff}&=&{\cal L}^{\rm bulk}_{\rm eff}
+{\cal L}^{\rm surf}_{\rm eff}
+\left(\partial_{\mu}\bar{\eta}\right)\left(\partial^{\mu}\eta\right)
+{\cal L}_{\rm gf} 
\;,
\eqa
where
\bqa
\nonumber
{\cal L}^{\rm bulk}_{\rm eff}&=&-{1\over4}F_{\mu\nu}F^{\mu\nu}+{c_1\over m^2}
F_{\mu\nu}n^{\alpha}\partial_{\alpha}n^{\beta}\partial_{\beta}F^{\mu\nu}
\\
&&
+{c_2\over m^2}F_{\mu\nu}n^{\mu}\partial^{\nu}
n^{\alpha}\partial^{\beta}F_{\alpha\beta}
+\delta{\cal L}^{}_{\rm bulk}
\;, \\
\label{surf}
{\cal L}^{\rm surf}_{\rm eff}&=&-{d_1\over4m}F_{\mu\nu}F^{\mu\nu}
\left[\delta(z)+\delta(z-L)\right]
+\delta{\cal L}^{}_{\rm surf}\;.
\eqa
The leading radiative correction 
$E_0^{(1)}$ is given by the surface term. 
It is calculated by using the photon propagator~(\ref{rtprop}). 
Comparing their  result with that of QED, they determine
$d_1=-3\alpha/32$.

We next consider the corrections to the Casimir free energy at finite
temperature. For convenience, we only consider the effective 
Lagrangian Eq.~(\ref{mell}), but the same results
are obtained using Eq.~(\ref{finnl}).

Finite temperature calculations are conveniently carried out 
in the imaginary
time formalism. 
The Euclidean Lagrangian corresponding to Eq.~(\ref{mell}) is
\bqa\nonumber
{\cal L}_{\rm eff}&=&
{1\over4}F_{\mu\nu}F_{\mu\nu}-b_1\left(n_{\mu}F_{\mu\alpha}\right)
\left(n_{\nu}F_{\nu\alpha}\right) 
\\&&
\label{mell2}
+\left(\partial_{\mu}\bar{\eta}\right)\left(\partial_{\mu}\eta\right)
+{\cal L}_{\rm gf}+\delta{\cal L}^{}_{\rm eff}
\;.
\eqa
In the following we use Feynman gauge, but 
results are independent of the gauge-fixing condition.
The propagator 
which satisfies the boundary conditions is
\bqa\nonumber
D^{S}_{\mu\nu}(x,x^{\prime})
&=&T
\sum_{n}\int{d^{d+1}k\over(2\pi)^{d+1}}
{\delta_{\mu\nu}\over K^2}
e^{-ik(x-x^{\prime})}\\ \nonumber
&&
-T\sum_{n}\int{d^{d}k_{\perp}\over(2\pi)^d}
{P_{\mu\nu}^{\perp}\over4\gamma\sinh\gamma L}
e^{-ik_{\perp}(x_{\perp}-x_{\perp}^{\prime})} \\
\nonumber
&& 
\times\left[e^{\gamma L}
\left(e^{-\gamma|z|}e^{-\gamma|z^{\prime}|}+
e^{-\gamma|z-L|}e^{-\gamma|z^{\prime}-L|}\right)\right. \\
\nonumber
&& 
\label{euclpr}
\left.
-\left(e^{-\gamma|z^{\prime}-L|}e^{-\gamma|z|}
+e^{-\gamma|z-L|}e^{-\gamma|z^{\prime}|}\right)
\right]\;,
\eqa
where $K^2=\omega_n^2+k^2$, $k_{\perp}=\sqrt{k_1^2+k_2^2}$,
$\gamma=\sqrt{\omega_n^2+k_1^2+k^2_2}$, and the projection
operator is
\bqa
P_{\mu\nu}^{\perp}=
\Bigg\{\begin{array}{cc}
\delta_{\mu\nu}-{k_{\mu}k_{\nu}\over k^2_{\perp}}& 
\hspace{-1.cm}{\rm for}\hspace{0.2cm}\mu, \nu\neq 3\\
0& \hspace{0.4cm}
{\rm for}\hspace{0.2cm}\mu=3\hspace{0.2cm}{\rm or}\hspace{0.2cm}\nu=3\;.
\end{array}
\eqa

The leading result ${\cal F}_0$ for the free energy is of course
reproduced by the effective theory, 
since the implementation of the boundary conditions 
is identical and independent of the presence of the fermion field.

The order-$\alpha$ correction ${\cal F}_1$ is given  by
\bqa
{\cal F}_{1}=-b_1\int_0^Ldz\;\langle F_{3\alpha}^2(z)\rangle\;.
\eqa
The first contribution comes from the $L$-independent part of the 
propagator
and is denoted by ${\cal F}_1^{(a)}$. It reads
\bqa\nonumber
{\cal F}_1^{(a)}&=&-b_1
dLT\sum_{n}\int{d^{d+1}k\over(2\pi)^{d+1}}{k_3^2\over K^2}
\\ 
&=&-b_1d
\pi^{d/2+1/2}
\zeta(-d-1)\Gamma\left(-\mbox{$d\over2$}-\mbox{$1\over2$}\right)LT^{d+2}
\label{lta}
\;.
\eqa
Taking the limit $d\longrightarrow2$, we obtain
\bqa
\label{ltab}
{\cal F}_1^{(a)}&=&-b_1{\pi^2L\over45\beta^4}\;.
\eqa
The second contribution arises from the $L$-dependent part of the
propagator and is denoted by ${\cal F}_1^{(b)}$. It reads
\bqa
\label{inter}
{\cal F}_1^{(b)}&=&
b_1
dLT\sum_n\int{d^{d}k_{\perp}\over(2\pi)^{d}}
\;{\gamma\over(e^{2\gamma L}-1)}\;.
\eqa
At low temperature, it is convenient to rewrite the exponential term
in Eq.~(\ref{inter})
using the formula
\bqa
\sum_m{1\over\gamma^2+k_m^2}&=&
{2L\over\gamma}\left[{1\over2}+{1\over e^{2\gamma L}-1}\right]
\;,
\eqa
where $k_m=m\pi/L$.
Summing over the Matsubara frequencies in Eq.~(\ref{inter}) yields
\bqa\nonumber
{\cal F}_1^{(b)}&=&
-{1\over2}b_1d\sum_{m}\int{d^{d}k_{\perp}\over(2\pi)^{d}}
\;{k_m^2\over\omega}\left[{1\over2}+{1\over e^{\beta\omega}-1}\right]
\\&&
\label{lttt}
+
b_1d
{\pi^{d/2+1}\zeta(-d-1)\Gamma
\left(
-\mbox{$d\over2$}-\mbox{$1\over2$}
\right)
\over\Gamma\left(\mbox{$1\over2$}\right)\beta^{d+2}}L
\;,
\eqa
where $\omega=\sqrt{k_{\perp}^2+k_m^2}$, and we have also integrated over 
$k_{\perp}$ in one of the terms in Eq.~(\ref{inter}).
The second term 
in Eq.~(\ref{lttt}) is exponentially suppressed, since the contribution
from $m=0$ vanishes.
Summing over $m$ in the first term and integrating over $k_{\perp}$, we obtain
\bqa\nonumber
{\cal F}_1^{(b)}&=&
-b_1d{\pi^{d/2+1}
\zeta(-d-1)\Gamma\left(\mbox{$1\over2$}-\mbox{$d\over2$}\right)
\over2^{d+1}\Gamma(\mbox{$1\over2$})L^{d+1}}
\\&&
\label{ltb}
+
b_1d
{\pi^{d/2+1}\zeta(-d-1)\Gamma
\left(
-\mbox{$d\over2$}-\mbox{$1\over2$}
\right)
\over\Gamma\left(\mbox{$1\over2$}\right)\beta^{d+2}}L
\;.
\eqa
Taking the limit $d\longrightarrow2$, Eq.~(\ref{ltb}) reduces to
\bqa
\label{ltbf}
{\cal F}_1^{(b)}&=&b_1{\pi^2\over240L^3}
+b_1{\pi^2L\over45\beta^4}\;.
\eqa
Adding 
Eqs.~(\ref{ltab}) and~(\ref{ltbf}), we obtain the complete 
order-$\alpha$ in result~(\ref{lt}).

At high temperature, the sum in Eq.~(\ref{inter})
is dominated by the static mode, and the
contributions from the nonzero Matsubara modes are exponentially
suppressed. In the high-temperature limit, ${\cal F}_1^{(b)}$
reduces to
\bqa\nonumber
{\cal F}_1^{(b)}&=&b_1dLT\int{d^dk_{\perp}\over(2\pi)^d}
{\gamma\over e^{2\gamma L}-1}\;, \\
&=&
b_1d{\zeta(d+1)\Gamma(d+1)\over2^{2d}\pi^{d/2}\Gamma(d/2)\beta L^d}\;.
\eqa
Taking the limit $d\longrightarrow2$, we obtain
\bqa
\label{htrad}
{\cal F}_1^{(b)}&=&b_1{\zeta(3)\over4\pi\beta L^2}\;.
\eqa
Adding Eqs.~(\ref{ltab}) and~(\ref{htrad}), we recover 
the order-$\alpha$ term in Eq.~(\ref{ht}).

In the high-temperature limit, the nonzero Matsubara modes decouple
and we can describe the system by a three-dimensional field theory
for the zero Matsubara modes by integrating out 
the nonzero Matsubara modes~\cite{gins,landsman,BN}.
The effective field theory consists of a scalar field $\bar{A}_0$
coupled to a three-dimensional gauge field 
$\bar{A}_i$ 
that (up to normalizations) can be identified with the zeroth modes
of the timelike and spatial components of the gauge field. 
In analogy with Eq.~(\ref{mell}), the effective three-dimensional field theory 
can be written as
\bqa\nonumber
{\cal L}_{\rm eff}&=&
{1\over2}(\partial_i\bar{A}_0)^2+{1\over4}F_{ij}^2
-e_1(n_i\partial_i\bar{A}_0)^2-e_2(n_iF_{ij})^2 \\
&&
\label{2d}
+\left(\partial_{i}\bar{\eta}\right)\left(\partial_{i}\eta\right)
+{\cal L}_{\rm gf}+\delta{\cal L}^{}_{\rm eff}
\;.
\eqa
The coefficients must be tuned so they reproduce static
correlators of full QED 
for 
$m^{-1}\ll\beta\ll L$. At leading order in the
fine structure constant, we have  $e_1=e_2=b_1$.
One term we omitted in Eq.~(\ref{2d}) is the coefficient of the unit
operator $f$~\cite{BN}
It is the term that arises from integrating out the
nonzero Matsubara modes in the graphs for the free energy.
At one loop, $f$ is directly given by contribution of the nonstatic
Matsubara modes to the free energy. It reads
\bqa
f&=&-{\pi^2L\over45\beta^3}\left[1-b_1\right]+{\zeta(3)\over2\pi\beta^2}
\label{f}
\;.
\eqa
The boundary conditions become $n_{i}\partial_iA_0=0 $ and 
$n_i\tilde{F}_{ij}=0$, where $n_i=(0,0,1)$.
The propagators are:
\bqa\nonumber
D^{S}(x,x^{\prime})&=&
\int{d^{d+1}k\over(2\pi)^{d+1}}
{1\over k^2}e^{-ik(x-x^{\prime})}\\ \nonumber
&&
-\int{d^dk_{\perp}\over(2\pi)^d}
{1\over4\gamma\sinh\gamma L}e^{-ik{\perp}
(x_{\perp}-x^{\prime}_{\perp})}\\
\nonumber
&& 
\times\left[e^{\gamma L}
\left(e^{-\gamma|z|}e^{-\gamma|z^{\prime}|}+
e^{-\gamma|z-L|}e^{-\gamma|z^{\prime}-L|}\right)\right. \\
\nonumber
&& 
\left.
-\left(e^{-\gamma|z^{\prime}-L|}e^{-\gamma|z|}
+e^{-\gamma|z-L|}e^{-\gamma|z^{\prime}|}\right)
\right]\;, 
\\ \nonumber
D_{ij}^{S}(x,x^{\prime})
&=&
\int{d^{d+1}k\over(2\pi)^{d+1}}
{\delta_{ij}\over k^2}
e^{-ik(x-x^{\prime})}
\\ \nonumber
&& \nonumber
-\int{d^dk_{\perp}\over(2\pi)^d}
{P_{ij}^{\perp}\over4\gamma\sinh\gamma L}e^{-ik{\perp}
(x_{\perp}-x^{\prime}_{\perp})}\\
\nonumber
&& 
\times\left[e^{\gamma L}
\left(e^{-\gamma|z|}e^{-\gamma|z^{\prime}|}+
e^{-\gamma|z-L|}e^{-\gamma|z^{\prime}-L|}\right)\right. \\
\nonumber
&& 
\left.
-\left(e^{-\gamma|z^{\prime}-L|}e^{-\gamma|z|}
+e^{-\gamma|z-L|}e^{-\gamma|z^{\prime}|}\right)
\right]\;,
\eqa
where $k^{\perp}_{\mu}=(k_1,k_2)$, 
$\gamma=\sqrt{k_1^2+k_2^2}$, $k^2=\gamma^2+k_3^2$,
and the projection
operator is now
\bqa
P_{ij}^{\perp}=
\Bigg\{\begin{array}{cc}
\delta_{ij}-{k_{i}k_{i}\over k^2_{\perp}}& 
\hspace{-1.cm}{\rm for}\hspace{0.2cm}i, j\neq 3\\
0& \hspace{0.4cm}
{\rm for}\hspace{0.2cm}i=3\hspace{0.2cm}{\rm or}\hspace{0.2cm}j=3\;.
\end{array}
\eqa
The implementation of the boundary conditions changes the one-loop free
energy in complete analogy with Eq.~(\ref{1lf}). 
Dropping terms that vanish with dimensional regularization, we have
\bqa\nonumber
{\cal F}_0&=&
{1\over2}d
\int{d^{d}k_{\perp}\over(2\pi)^{d}}
\log(1-{e^{-2\gamma L}})
\;.\\
&=&-
{\zeta(d+1)\Gamma(d)\over\pi^{d/2}2^{2d-1}\Gamma(d/2)L^d}\;.
\eqa
Taking the limit $d\longrightarrow2$, we obtain
\bqa
\label{l00}
{\cal F}_0&=&-
{\zeta(3)\over8\pi L^2}
\;.
\eqa
The order-$\alpha$ correction ${\cal F}_1$ is
\bqa
\label{3df}
{\cal F}_{1}=
\int_0^Ldz\;\Bigg[e_1\langle \bar{A}_0(z)\partial_z^2(z)\bar{A}_0\rangle
-e_2
\langle F_{3i}^2(z)\rangle\Bigg]\;.
\eqa 
Consider the first term in Eq.~(\ref{3df}):
\bqa\nonumber
\langle\bar{A}_0(z^{\prime})\partial_z^2\bar{A}_0(z)\rangle&=&
-\int{d^dk_{\perp}\over(2\pi)^d}{\gamma\over4\sinh kL}\Bigg[ \\ \nonumber
&& 
e^{\gamma L}\left[
e^{-\gamma(|z|+|z^{\prime}|)}
+e^{-\gamma(|z-L|+|z^{\prime}-L|)}
\right] \\
&&
-e^{-\gamma(|z|+|z^{\prime}-L|)}
-e^{-\gamma(|z-L|+|z^{\prime}|)}
\Bigg]\;.
\eqa
Setting $z=z^{\prime}$ and integrating over $z$, we obtain
\bqa\nonumber
\int_0^Ldz\;\langle\bar{A}_0(z)\partial_z^2\bar{A}_0(z)\rangle&=&
L\int{d^dk_{\perp}\over(2\pi)^d}{\gamma\over e^{2kL}-1} \\
&=&
\label{scc}
{\zeta(d+1)\Gamma(d+1)
\over2^{2d}\pi^{d/2}\Gamma(d/2)L^{d}}
\;.
\eqa
The contribution from the second term in Eq.~(\ref{3df}) 
is evaluated in the same manner and takes the value
\bqa
\int_0^Ldz\;
\langle F_{3i}^2(z)\rangle&=&(1-d)
{\zeta(d+1)\Gamma(d+1)
\over2^{2d}\pi^{d/2}\Gamma(d/2)L^{d}}\;.
\label{gc}
\eqa
Adding Eqs.~(\ref{scc}) and~(\ref{gc}), 
we obtain in the limit $d\longrightarrow2$
\bqa
{\cal F}_{1}=(e_1+e_2){\zeta(3)\over8\pi L^2}\;.
\label{eff}
\eqa
Adding Eqs.~(\ref{f}), (\ref{l00}),
and~(\ref{eff}), and multiplying by $T$ to get the
correct dimension, we recover the high-temperature limit~(\ref{ht}) of
the free energy obtained in QED.

In summary, we 
have shown that the effective field theories in Refs.~\cite{f+j,mel}
can reproduce the order-$\alpha$ correction to the Casimir free energy
at finite temperature in the same way as in the zero-temperature case.
Moreover, we have shown that the system undergoes dimensional reduction
at high temperature and can be described in terms of an
effective three-dimensional field theory.

The author would like to thank F. Ravndal for stimulating discussions.
This work was supported by the Stichting Fundamenteel Onderzoek der Materie
(FOM), which is supported by the Nederlandse Organisatie voor Wetenschapplijk
Onderzoek (NWO).

\end{multicols}

\end{document}